\documentclass[a4paper]{jpconf}

\usepackage[utf8]{inputenc}
\usepackage[T1]{fontenc}
\usepackage[english]{babel}

\usepackage{graphicx}
\usepackage{subfig}
\usepackage[usenames,dvipsnames]{color}
\usepackage{amsmath,amssymb}
\usepackage[final]{pdfpages}
\usepackage{geometry}

\begin{document}
\title{Multinucleation in the first-order phase transition of the 2d Potts model}

\author{F Corberi$^{1,2}$, L F Cugliandolo$^3$, M Esposito$^{2,3}$ and M Picco$^3$}
\address{$^1$ INFN, 
Gruppo Collegato di Salerno, Unit\`a di Salerno, Universit\`a  di Salerno, 
via Giovanni Paolo II 132, 84084 Fisciano (SA), Italy.}

\address{ $^2$ Dipartimento di Fisica ``E.~R. Caianiello'', Universit\`a  di Salerno, 
via Giovanni Paolo II 132, 84084 Fisciano (SA), Italy.}

\address{$^3$ Sorbonne Universit\'e, 
 Laboratoire de Physique Th\'eorique et Hautes Energies, CNRS UMR 7589,
4, Place Jussieu, Tour 13, 5\`eme \'etage,
75252 Paris Cedex 05, France.}

\ead{espoma@protonmail.ch}
\begin{abstract}
Using large-scale numerical simulations we studied the kinetics of the 2d $q$-Potts model for $q>4$ after a shallow subcritical quench from a high-temperature homogeneous configuration. This protocol drives the system across a first-order phase transition. The initial state is metastable after the quench and, for final temperatures close to the critical one, the system escapes from it via a multi-nucleation process. The ensuing relaxation towards equilibrium proceeds through coarsening with competition between the equivalent ground states. This process has been analyzed for different choices of the parameters such as the number of states and the final quench temperature. 
\end{abstract}
\section{Introduction}
Phase transitions are a widespread phenomenon in physics, mathematics and in nature in general \cite{Papon}. Some examples are melting ice, percolation, para- to ferro- transitions in magnetic systems, Bose-Einstein condensation, etc.

A system undergoes a phase transition when it changes its properties in a discontinuous way, that is, by exhibiting a discontinuity in a thermodynamic function. Specifically, we can make a distinction between two cases. When the first-order derivative of the Gibbs free energy $\mathcal{F}$ (i.e., the order parameter) is discontinuous, we say that a \emph{first-order} phase transition is taking place. When the discontinuity affects second- or higher-order derivatives of $\mathcal{F}$, we name them \emph{continuous} and we classify the former as \emph{second-order}.

In order to better understand the basic features of these two kinds of phase transitions, it is useful to consider the familiar Ising model, described by the Hamiltonian
\begin{equation}
H_I = - J \sum_{\langle ij \rangle} \sigma_i \sigma_j - H \sum_i \sigma_i 
\; ,
\end{equation}
where $J$ is a coupling constant, the first sum is restricted to nearest-neighbors on a lattice, $\sigma_i = \pm 1$ is a Boolean spin variable, 
and $H$ is an external magnetic field. The order parameter is the total magnetization 
\begin{equation}
M=\sum_i \sigma_i \; ,
\end{equation}
which is the first derivative of $\mathcal{F}$ with respect to the magnetic field.

It is well-known that the 2d Ising model undergoes a first-order phase transition at a fixed $T<T_c$ when the sign of the magnetic field is switched. At null field, the model exhibits a second-order phase transition when it is cooled from $T_i>T_c$ to $T_f<T_c$. 

\begin{figure}[!tbp]
 %  \begin{minipage}[b]{0.45\textwidth}
  \hspace{0.8cm} (a) \hspace{6cm} (b)
  \vspace{-0.7cm}
  \begin{center}
  \hspace{1cm}
    \includegraphics[width=0.37\textwidth]{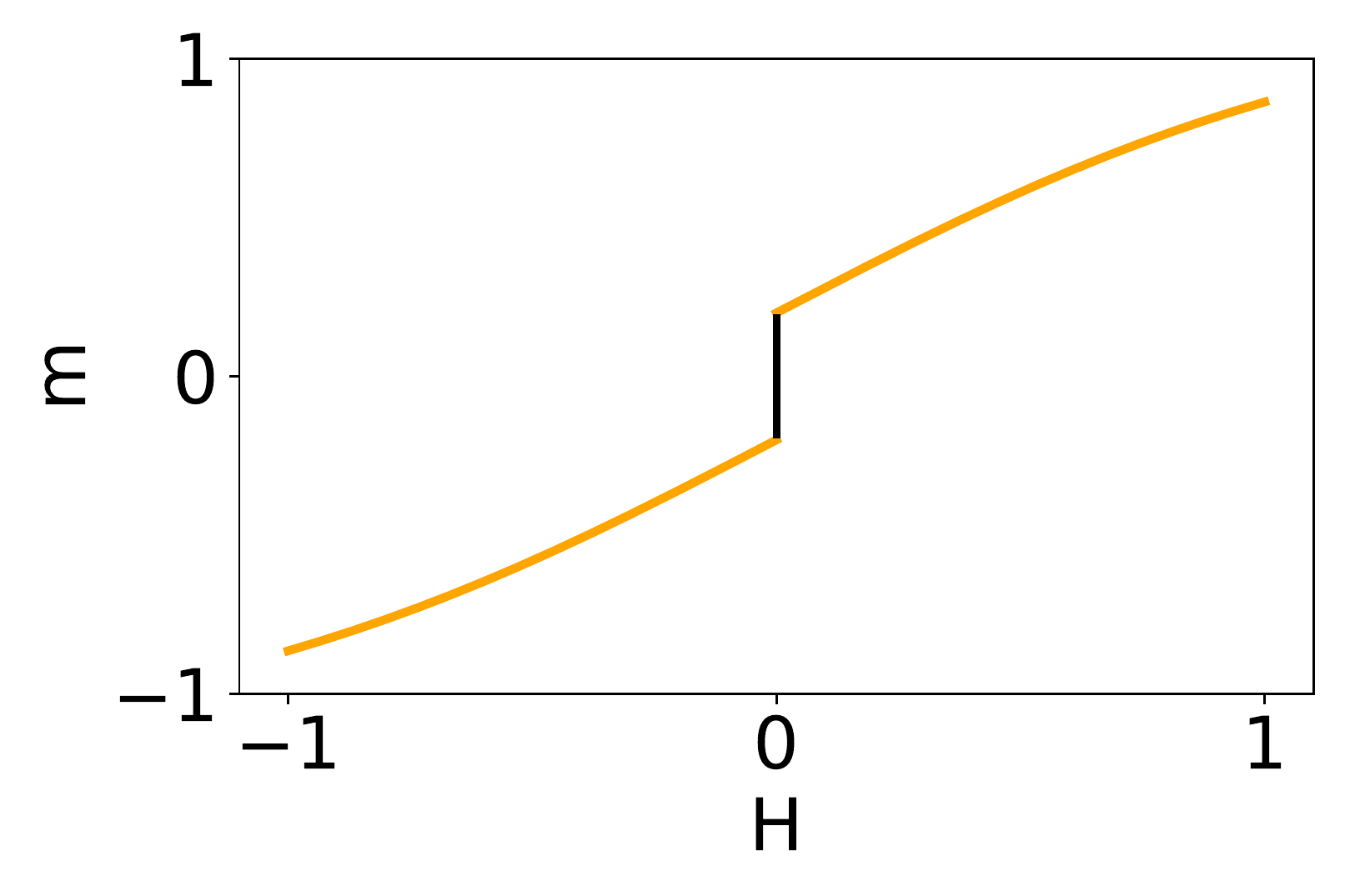}
%    \caption{Magnetization density, $m=M/N$, for $T<T_c$ in the first-order phase transition of the Ising model.}
%     \label{fig: magnFOPT}
%  \end{minipage}
%   \hspace{0.15cm}
%\hfill
%  \begin{minipage}[b]{0.45\textwidth}
\hspace{1cm}
    \includegraphics[width=0.37\textwidth]{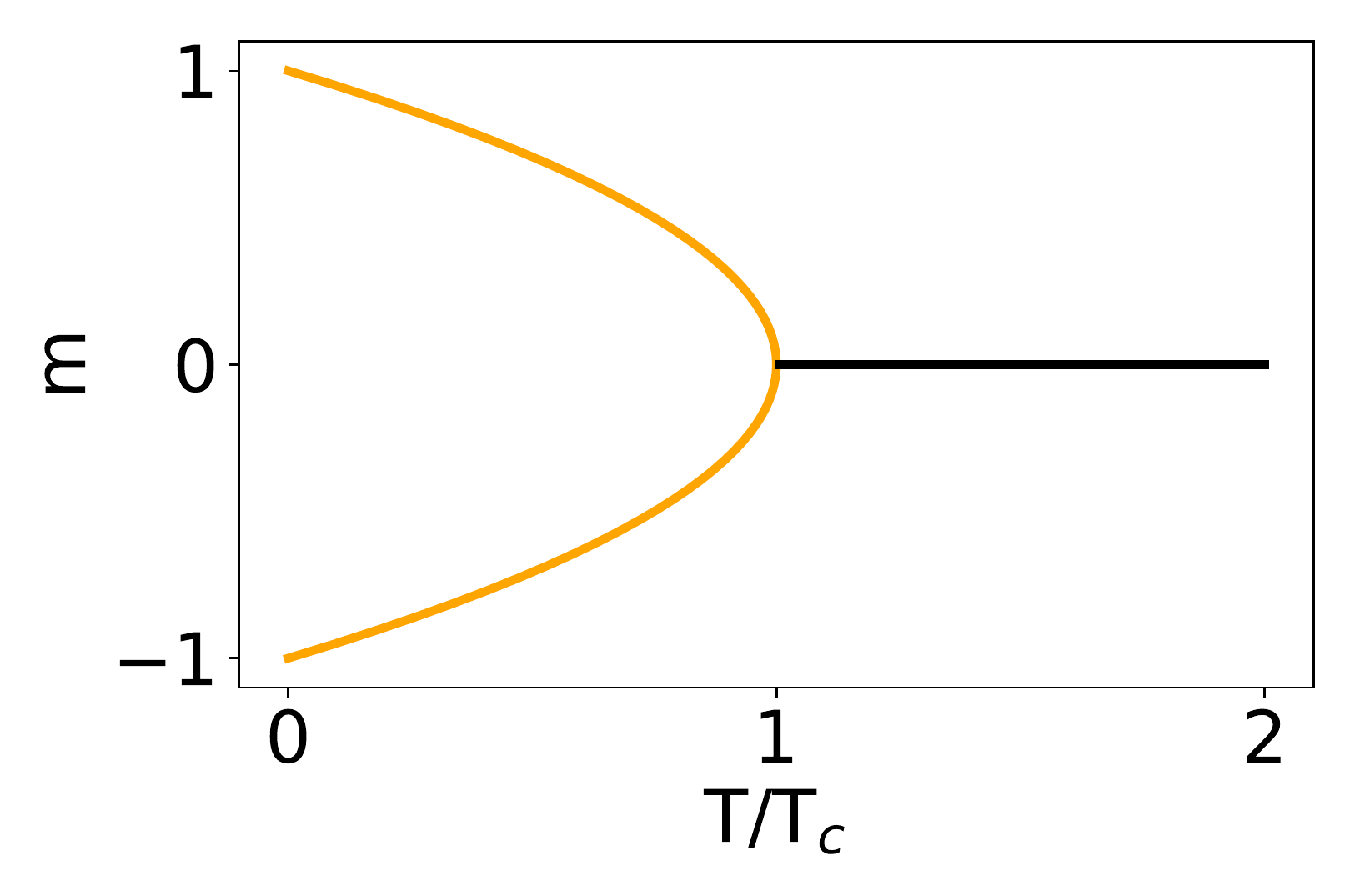}
    \end{center}
%    \caption{Magnetization density, $m=M/N$, in the second-order phase transition of the Ising model.}
   \caption{Magnetization density, $m=M/N$ of the Ising model, 
   (a) as a function of magnetic field at $T<T_c$   (first-order phase transition), 
   (b) as a function of temperature at zero magnetic field  (second-order phase transition).}
    \label{fig: magnSOPT}
%  \end{minipage}
\end{figure}

The magnetization (normalized by the number of spins) is plotted in Figs.~\ref{fig: magnSOPT}a and \ref{fig: magnSOPT}b for the two cases. It can be seen that $m$ is discontinuous at $H=0$ in the first-order phase transition and continuous in the second-order phase transition. In this last case, its first derivative -- that is, the magnetic susceptibility -- exhibits a discontinuity at $T=T_c$.

This is the overall picture when a control parameter is varied continuously and the system changes state via quasistatic transformations. If, on the other hand, the control parameter is varied abruptly, the system passes through non-equilibrium configurations until slow dynamical processes eventually lead it to equilibrium. 
Here we describe two of these processes that will be considered later, namely nucleation and coarsening. 
\begin{figure}
\begin{center}
    \includegraphics[width=1\textwidth]{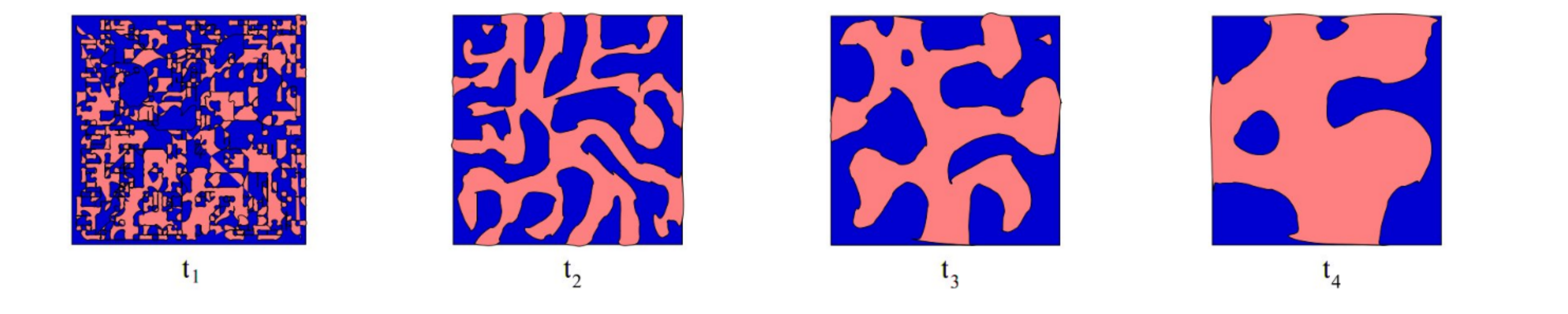}
    \end{center}
    \vspace{-0.3cm}
    \caption{Pictorial representation of a system during  coarsening at $t_1< \dots <t_4$.}
     \label{fig: coars}
\end{figure}
Coarsening characterizes the dynamics of  
second-order phase transitions~\cite{Bray,Puri,Corberi}, and it is the process whereby the 2d Ising model at $H=0$ orders after a quench from $T_i>T_c$ to $T_f<T_c$. In this case, it consists of the growth of domains driven by surface tension, as depicted in Fig. \ref{fig: coars}. Specifically, bigger domains devour smaller ones so that the typical domain size $R(t)$ grows in time as $R(t) \sim t^{1/2}$. Since (at finite temperature) equilibrium is reached when $R(t) \sim L$, relaxation takes an infinite time in the thermodynamic limit.

Nucleation characterizes the dynamics of 
first-order phase transitions~\cite{Langer1,Langer2,Langer3,Binder1,Binder2,Zeng_Oxtoby1}. It can take place in a system that is initially in a metastable state and has to exceed a free-energy barrier to reach equilibrium. The system tries to escape such metastable state via nuclei formation of the new phase into the old one. The contribution of a nucleus of size $r$ to the free energy consist of a positive surface term and a negative bulk term $\Delta \mathcal{F} = - \Delta f~r^d + s~r^{d-1}$, where $\Delta f$ is the bulk free energy, $s$ is the surface tension and $d$ is the dimensionality. Until a critical nucleus of size $r^* \propto s/\Delta f$ is created, the dynamics is inefficient as nuclei of a smaller size grow and shrink continuously without affecting the free energy. When the critical nucleus is created, the system has jumped over the free-energy barrier and small fluctuations start the relaxation towards the equilibrium state that can happen, for instance, via coarsening. The two processes, then, are not mutually exclusive, but can happen in the same system on different time scales.

The aim of this study is to characterize the dynamics following a sudden quench from initial conditions typical of equilibrium at $T_i = \infty$ to a temperature below the critical one such that the ordering process starts via a multinucleation of many competing phases and continues through coarsening. The simplest model that 
realises this phenomenology is the Potts model. We now proceed to define it and study it.

\section{Results}

The Potts model is described by the Hamiltonian \cite{Potts,Wu1,Baxter1}
\begin{equation}
H_P = - J \sum_{\langle ij \rangle} \delta(\sigma_i, \sigma_j)
\; ,
\end{equation}
where $J>0$ is a coupling constant, the sum is restricted to nearest-neighbors on a lattice, 
$\delta(a,b)$ is the Kronecker delta and $\sigma$ can take integer values from 1 to $q \ge 2$. This model is a generalization of the Ising model with zero external magnetic field, to which it reduces for $q=2$.

This model undergoes a phase transition at the critical temperature~\cite{Wu1,Baxter1}
\begin{equation}
T_c (q) = \frac{1}{\ln\left(1+\sqrt{q} \right)}
\; . 
\end{equation}
In particular, for $2 \le q \le 4$, the transition is of the second-order, while it is of the first-order for $q>4$.  The extent of the region of metastability can be 
determined with modern and powerful numerical methods~\cite{Ferrero2,Ferrero3}.

We defined the model on an $L \times L$ square lattice with periodic boundary conditions in the first-order phase transition regime, that is for $q>4$. We studied its kinetics after a quench from $T_i>T_c$ to $T_f<T_c$, for different values of the final temperature $T_f$, the number of available states $q$ and the system size $L$. Starting from a completely disordered configuration -- corresponding to an infinite-temperature state -- the dynamical evolution proceeds via a Monte Carlo Markov Chain \cite{Newman1}, where a lattice spin and a number $l \in [1,q\,]$ are randomly picked and the spin is changed from state $r$ to state $l$ with the Metropolis transition rate
\begin{equation}
w_{rl} = \min \left(1, e^{-\beta \Delta E_{rl}} \right) \; ,
\end{equation}
where $\beta=T_f^{-1}$ (we set the Boltzmann constant to unity) 
and $\Delta E_{rl}$ is the energy difference between the configurations before and after the attempted spin flip.
A Monte Carlo Step (MCS) is the time unit and corresponds to $N=L \times L$ spin flip attempts.

The dynamical evolution of the Potts model has been studied in previous works that were mainly focused on the coarsening regime \cite{Ferrero_Cannas1,Ferrero_Cannas2,Leyvraz1,Cugliandolo_Coarsening1,Cugliandolo_Coarsening2,Krapivsky}, (see discussion below). 
We directed instead our analysis to the nucleation process.

Our primary quantity of interest is the excess of energy density~\cite{Ferrero_Cannas1}
\begin{equation}
\phi_E (t) = e(t) - e(\infty) \; ,
\end{equation}
where $e(t)$ is the energy density at a time $t$ after the quench and $e(\infty)$ is the equilibrium energy density. The latter can be computed starting from a completely ordered configuration, that is to say, all spins talking the same value among the $q$ possible ones (equilibrium at zero temperature) 
and letting the system relax towards equilibrium at the desired temperature. 

\begin{figure}[!tbp]
% \centering
 \begin{minipage}[b]{1\textwidth}
%\hspace{1.5cm} (a) \hspace{5.cm} (b)
\vspace{-1.5cm}
   \begin{center}
    \includegraphics[width=0.44\textwidth]{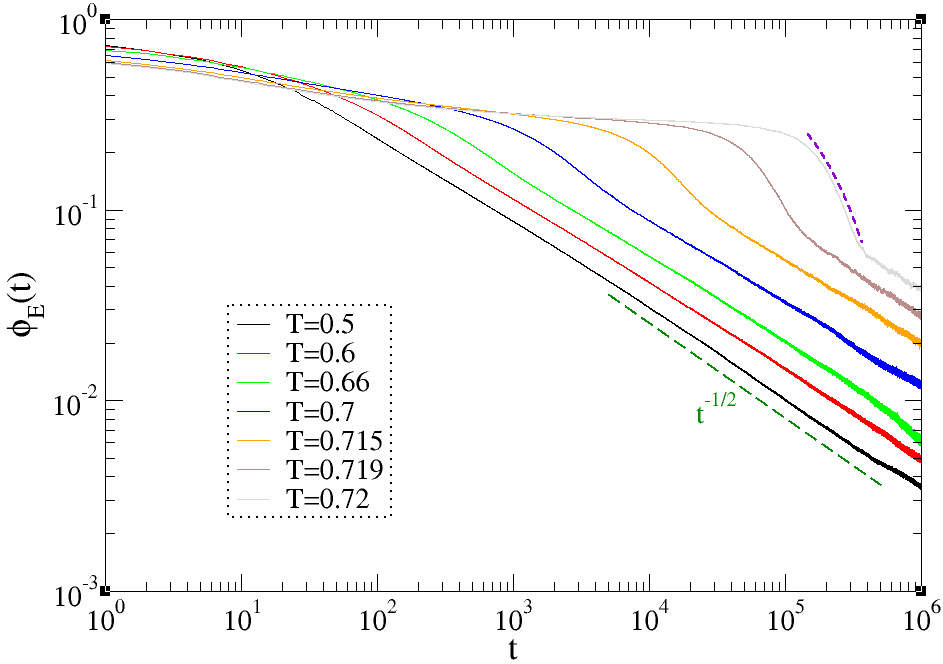}
%     \label{fig: inset}
%  \end{minipage}
\hspace{0.5cm}
%\hfill
%  \begin{minipage}[b]{0.48\textwidth}
    \includegraphics[width=0.44\textwidth]{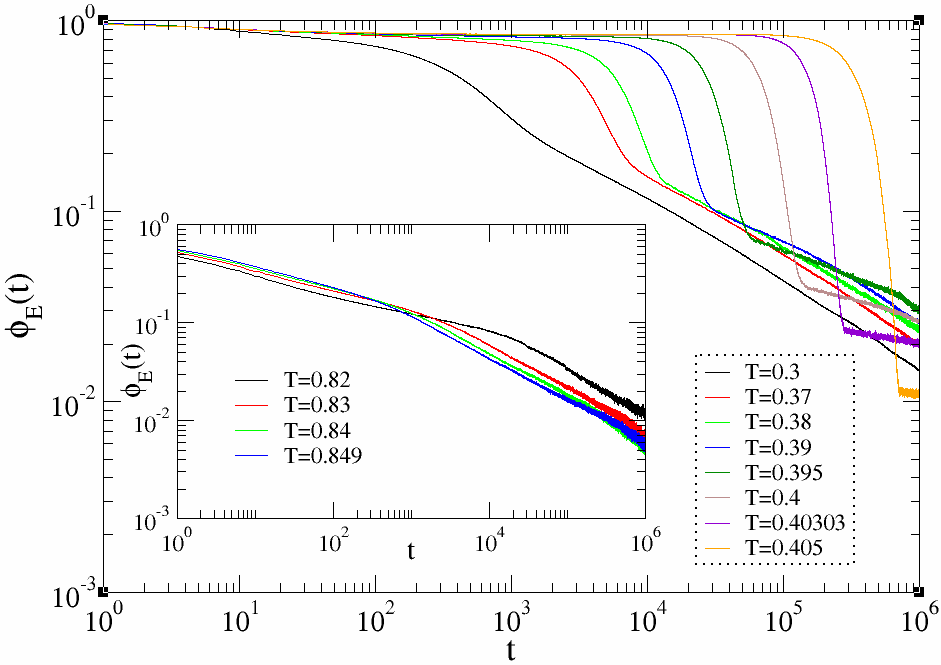}
    \end{center}   
    \caption{$\phi_E$ vs $t$ for $q=9$ ($T_c=0.71235$), $L=1000$ and different values of the temperature 
    $T_f$ that are given in the key ({\itshape Left Panel}). In the right panel, $\phi_E$ vs $t$ for $q=100$ ($T_c=0.41703$), $L=700$ and different values of $T_f$ also in the key. In the inset, the same but for $q=5$ ($T_c=0.85153$).}
    \label{fig: 100qp}
 \end{minipage}
\end{figure}

The behavior of the excess energy $\phi_E$ is plotted in Figs.~\ref{fig: 100qp}a and \ref{fig: 100qp}b for $q=5, 9$ and $100$, for different sizes $L$ and values of the final temperature $T_f$. First of all, it can be seen that for $T \lesssim T_c$, we can identify three different regimes. After a very short transient, 
$\phi_E$ reaches a long-lasting plateau. At a time $\tau_N (T_f)$ -- which we call \emph{nucleation time} -- $\phi_E$ decreases abruptly to a value that depends on the final temperature $T_f$. This is consistent with the picture of nucleation given before, where the system is initially stuck in the metastable state and tries to escape via nuclei formation. The dynamics are inefficient until several critical nuclei are created around a time $\tau_N (T_f)$, which results in a constant value of $\phi_E$. At $t=\tau_N$, the system exceeds the free-energy barrier. In Fig.~\ref{fig: 100qp}a, the violet dashed curve represents the function $f(t) = 0.51\exp(-5.98 \cdot 10^{-6}~t)$, which is the best exponential fit for the fast decay at $T_f=0.72$.

Relaxation well after $\tau_N(T_f)$ proceeds via coarsening, as can be seen by looking at the power-law behavior of the excess of energy $\phi_E \propto \lambda_q(T_f) \, t^{-1/2}$, which implies \cite{Bray} $R(t) \propto t^{1/2}$, as expected in coarsening.  Notice how the first-order phase transition is less evident for $q=5$ (inset of Fig.~\ref{fig: 100qp}a). This is probably due to the proximity to the case $q=4$, where the transition turns second-order. 

\begin{figure}[h!]
\centering
\vspace{-0.5cm}
{\includegraphics[width=.48\textwidth]{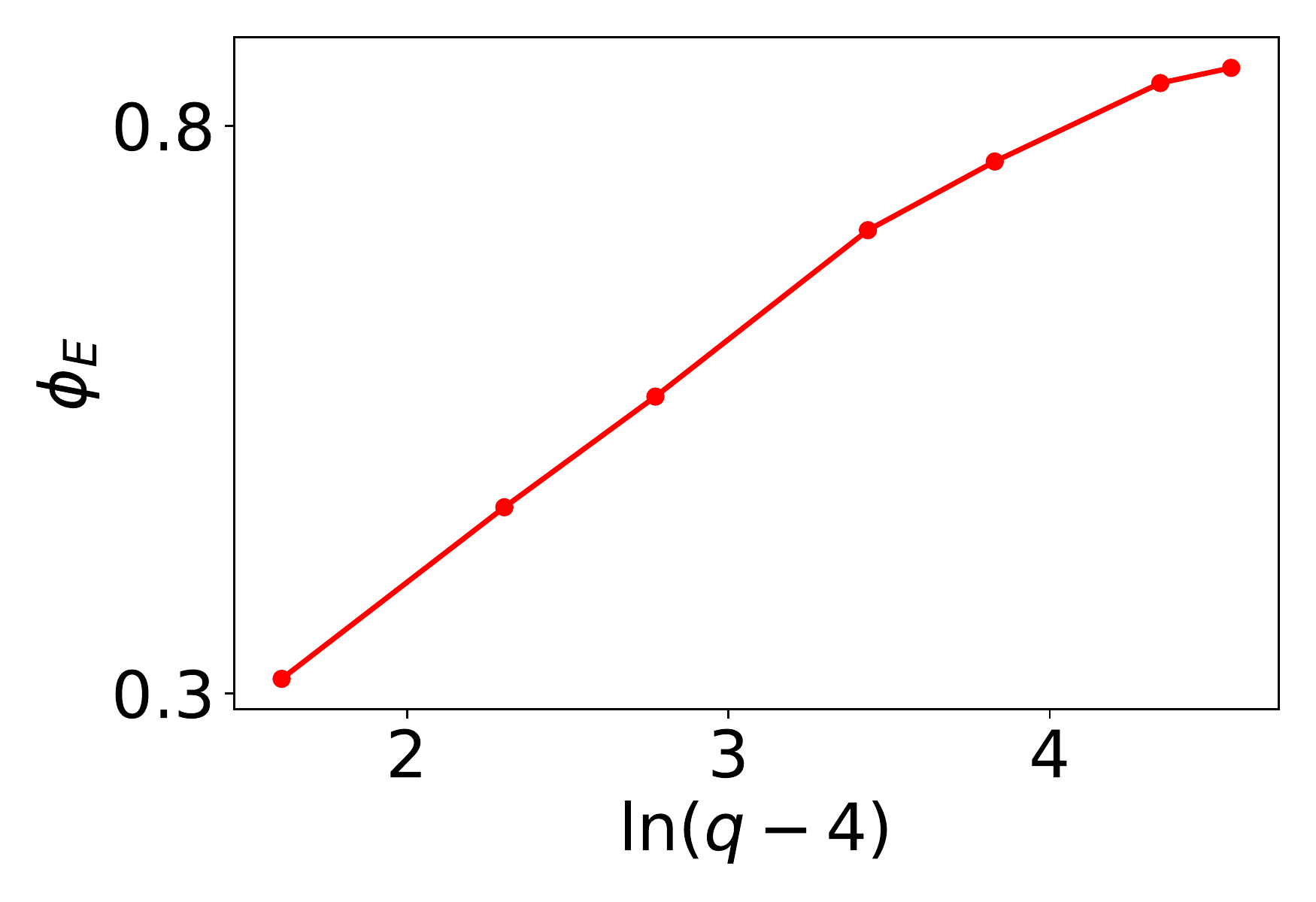}}
\caption{Value of $\phi_E$ on the plateau as a function of the logarithm of $q-4$.}
\label{fig: phipl}
\end{figure}

It is evident that both the shape of the fast decay of $\phi_E$ after $\tau_N$ and the temperature range in which nucleation dominates the  dynamics depend on the value of $q$. Specifically, the rapid relaxation becomes sharper and is observed for a wider temperature range the greater the value of~$q$. Let us notice that the value of the excess of energy in correspondence of the plateau $\phi_E^*$ depends on $q$, but it is rather temperature-independent. We have plotted in Fig. \ref{fig: phipl} the absolute value of $\phi_E^*$ as a function of the number of states. For $q<50$, the value of $\phi^*_E$ is well approximated by the logarithm of $q-4$. For greater values of $q$ one has saturation to a constant value $\phi_E^* \simeq 0.85$. 

\begin{figure}[b!]
\centering
\subfloat[][\emph{\label{fig: ss1}}]
{\includegraphics[width=.17\textwidth]{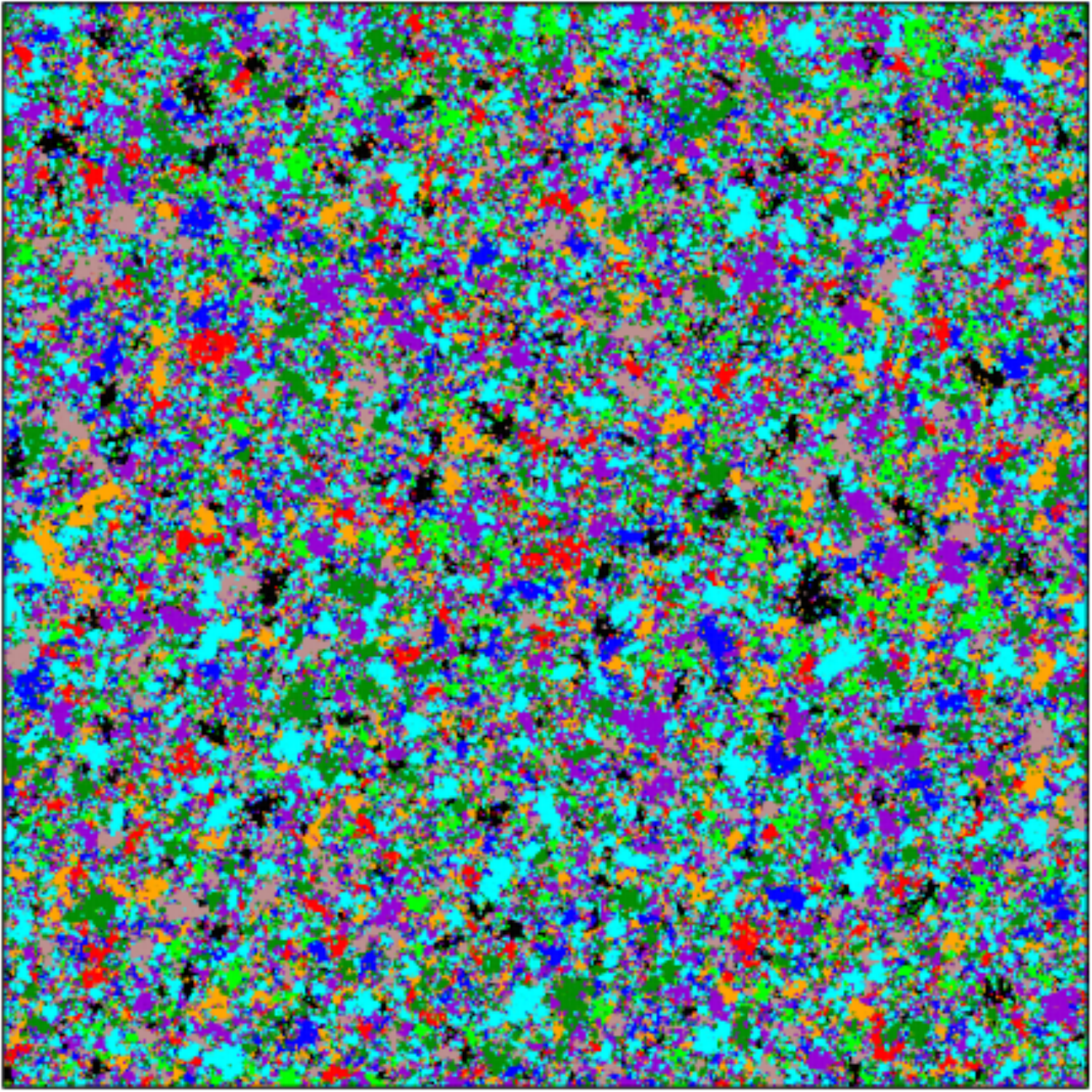}} \;
\subfloat[][\emph{\label{fig: ss2}}]
{\includegraphics[width=.17\textwidth]{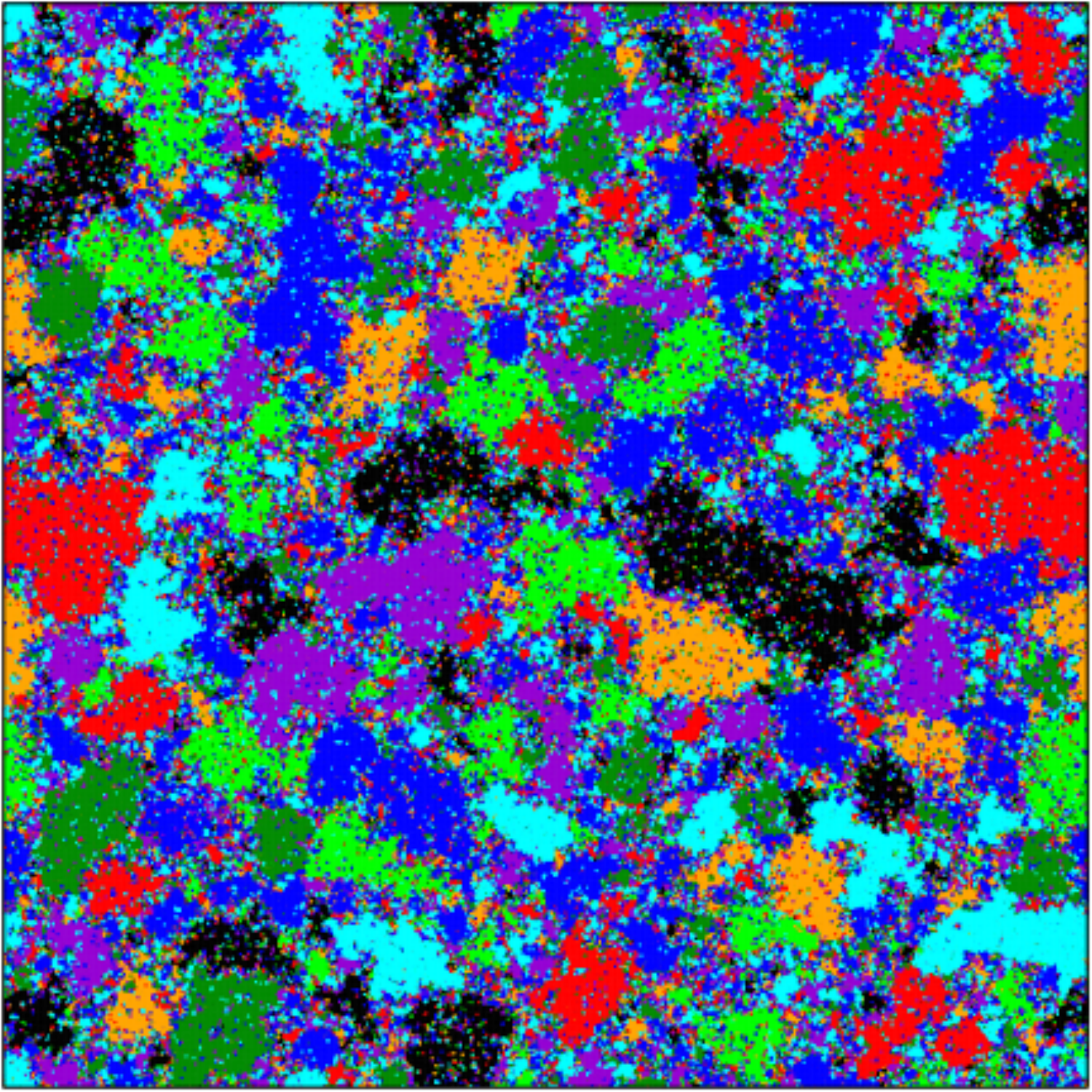}} \;
\subfloat[][\emph{\label{fig: ss3}}]
{\includegraphics[width=.17\textwidth]{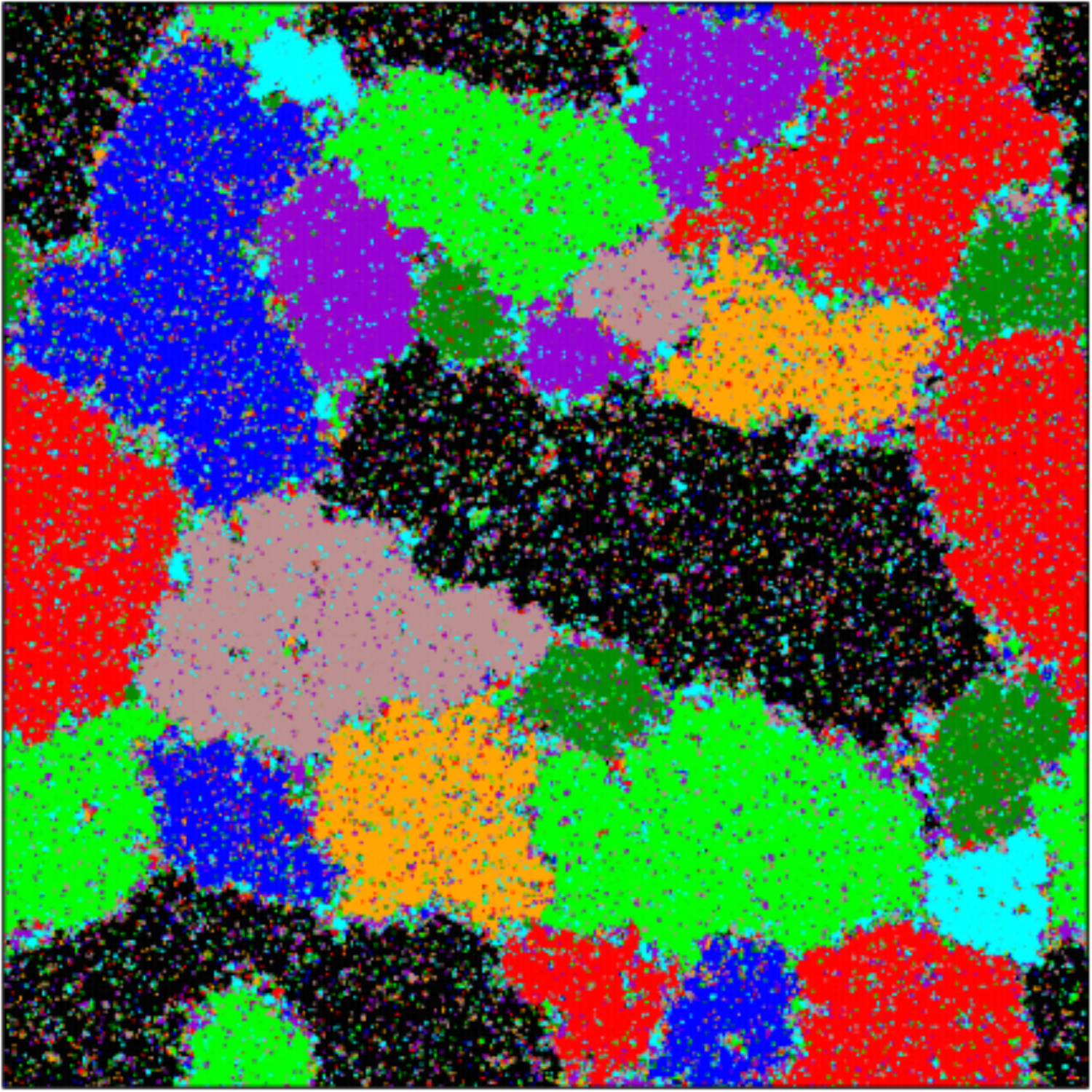}} \; 
\subfloat[][\emph{\label{fig: ss4}}]
{\includegraphics[width=.17\textwidth]{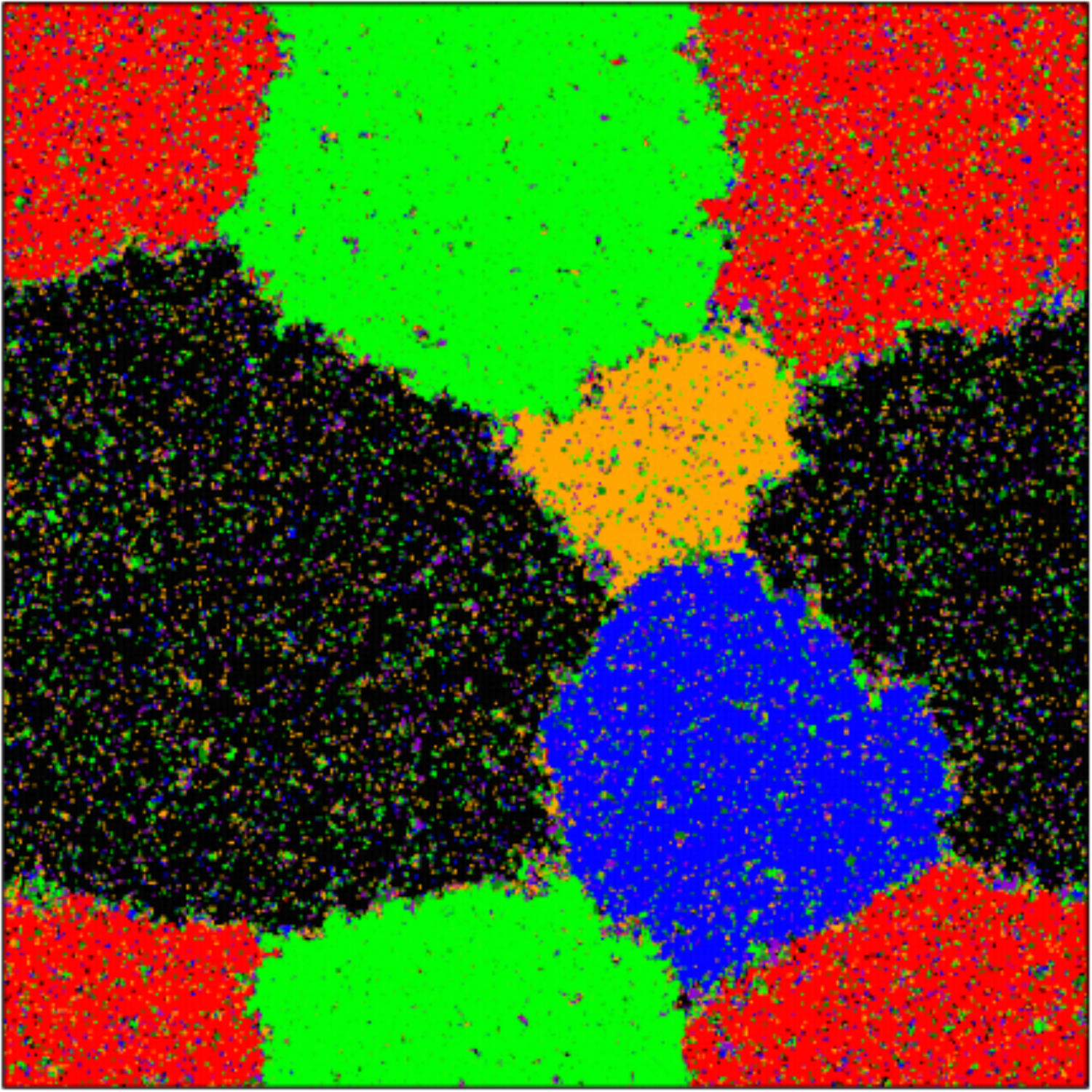}} 
\caption{Snapshots of the lattice at different times for $q=9$, $L=1000$ and $T=0.715$ (orange curve in Fig.~\ref{fig: 100qp}a) in the three stages of plateau (Fig.~\ref{fig: ss1}), fast decay (Fig.~\ref{fig: ss2}) and coarsening (Fig.~\ref{fig: ss3} and Fig.~\ref{fig: ss4}). Notice that -- at $t=10^6$ -- four phases have been eliminated.}
\label{fig: ss}
\end{figure}

Four snapshots of the lattice for $q=9$, $L=1000$ and $T=0.715$ are shown in Fig. \ref{fig: ss} for the three different regimes of $\phi_E$. At $t=10^3$ -- in the plateau region -- the system is still in a completely disordered configuration, qualitatively similar to the one at $t=0$, a sign that the dynamics is initially inefficient. Some larger domains are visible but they are too small to trigger nucleation. During the fast decay,  
nuclei of different phases grow very rapidly, see Fig. \ref{fig: ss2}. Coarsening, as shown in Figs.~\ref{fig: ss3} and \ref{fig: ss4} is characterized by growth of bigger domains to the detriment of smaller ones. At $t=10^6$, the time at which the snapshot in Fig. \ref{fig: ss4}  was recorded, four phases are no longer present in the system.

\section{Conclusions}
In this work we investigated the dynamics of the Potts model for $q>4$ after a quench from $T_i>T_c$ to $T_f<T_c$. Relaxation proceeds via multinucleation for $T_f \lesssim T_c$, followed by coarsening. This behavior has been characterized by looking at the excess of energy, that shows a temperature-independent plateau at the beginning, followed by a fast decay -- which is best fitted with an exponential -- and ends with a coarsening-like behavior $\phi_E \propto t^{-1/2}$. The behavior of the plateau energy $\phi_E^*(q)$ is consistent with the logarithm of $q-4$. During coarsening, phases are eliminated. In the light of our results, it can be interesting to investigate the spontaneous symmetry breaking mechanism whereby this elimination takes place, and compare it with the coarsening process that characterizes the dynamics of the Ising model, to identify possible similarities and differences. It could also be useful to study the geometrical properties of the clusters to understand whether their evolution is consistent with the Neumann-Mullins law \cite{Neumann, Mullins}.

\vspace{0.5cm}

\noindent
{\bf Acknowledgements}
LFC wishes to thank F. Krzakala for early discussions on this problem. She is a member of Institut Universitaire de France.

\end{document}